\documentclass[letter]{aa}
\usepackage[english]{babel}
\usepackage{graphicx}
\usepackage{amsmath}
\usepackage{amssymb}
\usepackage[]{natbib}
\bibpunct{(}{)}{;}{a}{}{,}

\def\apj{ApJ}%
\def\apjl{ApJ}%
\def\apjs{ApJS}%
\def\ao{Appl.~Opt.}%
\def\aap{A\&A}%
\def\aaps{A\&AS}%
\def\mnras{MNRAS}%
\def\nat{Nature}%

\begin{document}
	 \title{Hot circumstellar material resolved around \object{$\beta$~Pic} with VLTI/PIONIER\thanks{Based on data collected at the ESO Paranal Observatory under commissioning time and programme 088.C-0266.}}
   \titlerunning{Hot circumstellar material resolved resolved around \object{$\beta$~Pic} with VLTI/PIONIER}
   %\subtitle{}

   \author{D. Defr\`ere\inst{1}
           \and
          J. Lebreton\inst{2}
          \and
          J.-B. Le Bouquin\inst{2}
          \and
          A.-M. Lagrange\inst{2}
          \and
          O. Absil\inst{3,}\thanks{FNRS Postdoctoral Researcher.}
          \and
          J.-C. Augereau\inst{2}
          \and
          J.-P. Berger\inst{4}
          \and
          E. di Folco\inst{5}
          \and
          S. Ertel\inst{2}
          \and
          J. Kluska\inst{2}
          \and
          G. Montagnier\inst{4}
          \and
          R. Millan-Gabet\inst{6}
          \and
          W. Traub\inst{7}
          \and
          G. Zins\inst{2}
         }

   %\offprints{D. Defr\`ere}

   \institute{Max Planck Institut f\"{u}r Radioastronomie, Auf dem H\"{u}gel 69, 53121 Bonn, Germany\\
              \email{ddefrere@mpifr-bonn.mpg.de}
         \and
         		 UJF-Grenoble 1 / CNRS-INSU, Institut de Plan\'etologie et d'Astrophysique de Grenoble (IPAG) UMR 5274, France
         \and
             Universit\'e de Li\`ege, 17 All\'ee du Six Ao\^ut, B-4000 Li\`ege, Belgium
         \and
             European Southern Observatory, Alonso de Cordova, 3107, Vitacura, Chile
         \and
         		 Universit\'e de Bordeaux, Observatoire Aquitain des Sciences de l'Univers  UMR 5804, Floirac, France
         \and
         		 NASA Exoplanet Science Institute (Caltech), MS 100-22, 770 South Wilson Avenue, Pasadena, CA 91125, USA
          \and
             Jet Propulsion Laboratory (NASA/JPL), MS 301-355, 4800 Oak Grove Drive, Pasadena, CA 91109, USA
             }

   \date{Received; accepted}

  \abstract
   {}
   {We aim at resolving the circumstellar environment around \object{\object{$\beta$~Pic}} in the near-infrared in order to study the inner planetary system ($<200$~mas, i.e., $\sim$4~AU).}
   {Precise interferometric fringe visibility measurements were obtained over seven spectral channels dispersed across the H band with the four-telescope VLTI/PIONIER interferometer. Thorough analysis of interferometric data was performed to measure the stellar angular diameter and to search for circumstellar material.}
   {We detected near-infrared circumstellar emission around \object{$\beta$~Pic} that accounts for 1.37\% $\pm$ 0.16\% of the near-infrared stellar flux and that is located within the field-of-view of PIONIER (i.e., $\sim$200\,mas in radius). The flux ratio between this excess and the photosphere emission is shown to be stable over a period of 1 year and to vary only weakly across the H band, suggesting that the source is either very hot ($\gtrsim1500$~K) or dominated by the scattering of the stellar flux. In addition, we derive the limb-darkened angular diameter of \object{$\beta$~Pic} with an unprecedented accuracy ($\theta_{\rm LD}$= 0.736 $\pm$ 0.019~mas).}
   {The presence of a small H-band excess originating in the vicinity of \object{$\beta$~Pic} is revealed for the first time thanks to the high-precision visibilities enabled by VLTI/PIONIER. This excess emission is likely due to the scattering of stellar light by circumstellar dust and/or the thermal emission from a yet unknown population of hot dust, although hot gas emitting in the continuum cannot be firmly excluded.}


   \maketitle

\section{Introduction}

The young \citep[$\sim$12$^{+8}_{-4}$~Myr,][]{Zuckerman:2001} A6V-type star $\beta$\,Pic (HD~39060, A6V, 19.3\,pc) is surrounded by a famous planetary system, which is a prime target for understanding planetary system formation and evolution. Since its discovery \citep{Smith:1984}, successive generations of telescopes have reported the detection of an edge-on debris disc with several distinctive features suggestive of a multiple-belt system \citep[][]{Telesco:2005}, star-grazing comets \citep[``falling evaporating bodies'',][]{Beust:1990}, circumstellar gas \citep[e.g.,][]{Hobbs:1985, Roberge:2006}, and a 9-M$_{\rm Jup}$ planetary companion orbiting at a projected distance of approximately 4.3\,AU \citep[][]{Lagrange:2009}. The existence of other planets seems likely \citep[][]{Freistetter:2007} and might explain several asymmetries identified in the debris disc, including a warp at $\sim$50~AU \citep{Mouillet:1997,Augereau:2001} inclined by $\sim$4$^\circ$ with respect to the outer disc \citep{Lagrange:2012a}.

Over the past few years, the close environment ($\lesssim$ a few AU) of $\beta$\,Pic has been the focus of several studies trying to detect a putative sub-stellar companion \citep[e.g.,][]{Absil:2010, Lagrange:2012b}. In this paper, we present high-accuracy H-band interferometric observations of \object{$\beta$~Pic} that were originally aimed at a deep search for faint companions in the inner disc but which revealed instead the presence of a more extended circumstellar emission. Using our spectrally dispersed measurements and revisiting archival K-band VLTI/VINCI measurements, we constrain the nature of this excess emission and discuss briefly its origin.

\section{Observations and data reduction}

\begin{figure}[!t]
\centering
\vspace{0.1cm}
\includegraphics[height=7cm]{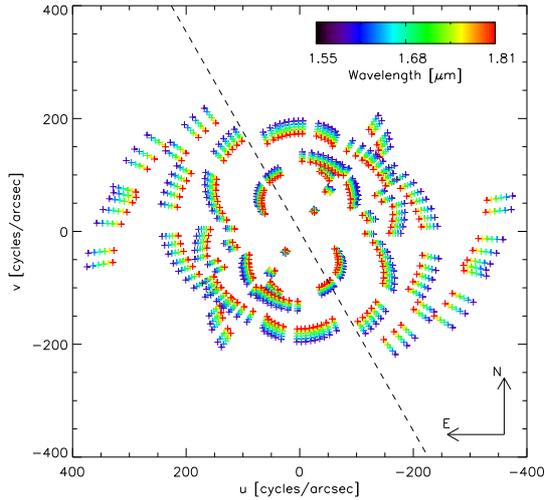}
\caption{Sampling of the Fourier (\textit{u,v}) plane obtained for the complete data set (one colour per spectral channel). The orientation of the outer disc midplane \citep[i.e., $29\fdg5$][]{Boccaletti:2009} is represented by the black dashed line.}
\label{fig:uv-plane}
\end{figure}

Interferometric data were obtained with the PIONIER combiner \citep[][]{LeBouquin:2011} at the VLT interferometer on three different occasions (see Table~\ref{tab:data_overview}). The fringe measurements were dispersed over seven spectral channels across the H band (1.50 - 1.80\,$\mu$m) but in the following we discard the first spectral channel (1.50 - 1.55~$\mu$m) that is systematically of lower quality and not suitable for the high-precision visibility measurements required by our programme. Observations of $\beta$\,Pic were interleaved with observations of reference stars to calibrate the instrumental contribution in the observed quantities. Calibrators were chosen close to $\beta$\,Pic, in terms of both position and magnitude, from the catalogue of \cite{Merand:2005}. The total \textit{u-v} plane covered by the observations is shown in Fig.~\ref{fig:uv-plane}.

Data were reduced and calibrated with the \texttt{pndrs} package \citep[][]{LeBouquin:2011}. We focus here on the squared visibilities ($\mathcal{V}^2$) to measure both the stellar angular diameter and search for circumstellar material. The final calibrated data set ($\mathcal{V}^2$) is shown in Fig.~\ref{fig:globalfit}. The search for faint companions by means of a closure-phase analysis will be presented elsewhere (Lagrange et al.\ in prep).
%and analysed with the \texttt{absilfit} package \citep[e.g.,][]{Absil:2006, Absil:2008b, Absil:2009, Defrere:2011}
\section{Data analysis}

\begin{table}[!t]
	\begin{center}
		\caption{Overview of the data obtained with VLTI/PIONIER on $\beta$ Pic.}\label{tab:data_overview}
	 \begin{tabular}{c c c c c c}
	   \hline
	   \hline
			ID & Date & Config. & Seeing & OB$^a$ & Calib.$^b$\\
	   \hline
				%A & 2010/11/29 & E0-G0-H0-I1 & 8  & 3,4\\ (3) HD\,39148 (K4III), (4) HD\,39312 (K1III),
				A & 2010/12/04 & E0-G0-H0-I1 & 0.9'' & 2  & 3,4,5\\
				B & 2010/12/20 & A0-G1-I1-K0 & 1.0'' & 5  & 1,2,3 \\
				C & 2011/11/02 & D0-G1-H0-I1 & 0.9'' & 14 & 2,3,4 \\
			\hline
		\end{tabular}\\
	\end{center}
{\scriptsize$^a$ An observing block consists in a single observation \citep[see more details in][]{LeBouquin:2011}}.\\
{\scriptsize$^b$ Calibrator stars correspond to (1) \object{HD\,34642} (K1IV), (2) \object{HD\,35765} (K1III),  (3) \object{HD\,39640} (G8III), (4) \object{HD\,46365} (K3III), and (5) \object{HD\,223825} (G9III).}
\end{table}

The calibrated $\mathcal{V}^2$ were fitted to a range of models consisting of an oblate limb-darkened photosphere surrounded by a uniform emission (``disc'') filling the entire field-of-view of PIONIER on the auxiliary telescopes. Under typical seeing conditions, this field-of-view can be approximated by a Gaussian profile with a full width at half maximum of 400~mas \citep[][]{Absil:2011}, equivalent to 4~AU in radius at the distance of \object{$\beta$~Pic}. Our model is based on two free parameters, namely the limb-darkened angular diameter of the star and the disc/star contrast. The distortion of the photosphere produced by the rapid rotation \citep[\textit{v}~$\sin i$ = 130\,km/s,][]{Royer:2007} was considered to produce a realistic model of the star. Following the parametric approach of \cite{Absil:2008b}, the distorted photosphere was modeled by an ellipse with an oblateness of 1.038 and a rotation axis perpendicular to the outer disc midplane \citep[which has a position angle of $29\fdg5$,][]{Boccaletti:2009}. The $\mathcal{V}^2$ expected from the limb-darkened photosphere was then estimated according to \cite{Hanbury:1974} considering a linear limb-darkening H-band coefficient of 0.24 \citep[][]{Claret:1995}.

\begin{figure}[!t]
\centering
\includegraphics[height=7.05cm]{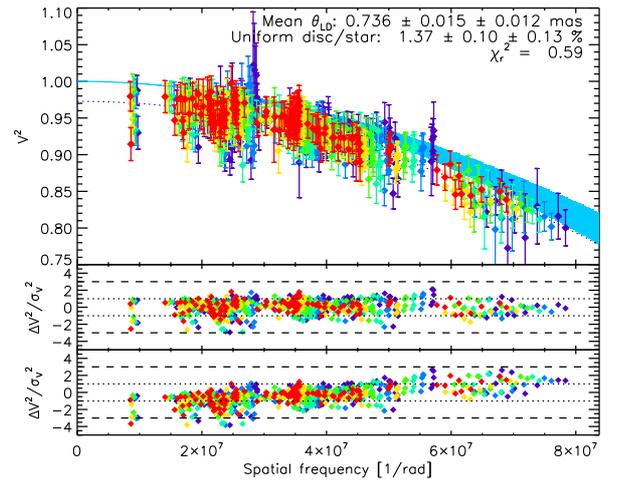}
\caption{Expected squared visibility of the limb darkened photosphere (blue solid line) as a function of the spatial frequency, along with the measured squared visibilities and related $1-\sigma$ error bars (one colour per wavelength). The thickness of the blue solid line corresponds to the 3-$\sigma$ error related to the uncertainty on the stellar diameter. The best-fit model is represented by the dotted blue line with the residuals of the fit given in the middle panel. It corresponds to a limb-darkened photosphere of 0.736 $\pm$ 0.015 $\pm$ 0.012~mas in diameter surrounded by a uniform circumstellar emission of 1.37 $\pm$ 0.10 $\pm$ 0.13\% in the H band. The bottom panel gives the residuals obtained by fitting only the stellar diameter (no circumstellar emission).}\label{fig:globalfit}
\end{figure}

To search for circumstellar material, we compared the measurements to the expected $\mathcal{V}^2$ of the stellar photosphere. The circumstellar emission then appears as a $\mathcal{V}^2$ deficit at short baseline lengths as detailed in \cite{Absil:2006b}. In this first approach, we assume that the disc/star contrast does not depend on the wavelength and fit all measurements simultaneously with this single parameter. We used the bootstrapping method to compute the statistical error bars on the stellar angular diameter and the disc/star contrast. Considering correlation between the baselines, the spectral channels, and the successive measurements of the same baseline, we found that correlations mostly occur between all spectral channels of the same measurement. Hence, we performed the bootstrapping by drawing all spectral channels together, which corresponds to 126 independent data sets out of our $6\times6\times21=726$ visibility measurements. We randomly produced 100 data sets from the original data and fitted them separately to our model. The standard deviation on the derived best-fit parameters then gave us reliable error bars. Finally, we considered an additional error to take the chromatic behaviour of the beam combiner and the different colours between \object{$\beta$~Pic} (A6V) and its calibrators (G8III to K4III) into account. Based on the spectral shape of the transfer function across the H band (maximum variation of 40\% over the entire band), we derived a maximum systematic error on the $\mathcal{V}^2$ of $\pm$0.2\% in a single spectral channel \citep[see more details in][]{Defrere:2011}. This value must be considered as very conservative since all other sources of systematic error are expected to be much smaller thanks to our optimised observing strategy, i.e. spectrally-dispersed observations obtained at various epochs on various configurations with different calibrators that are close to the science target in both magnitude and position.
\begin{figure}[!t]
	\begin{flushleft}
		\includegraphics[height=6.5 cm]{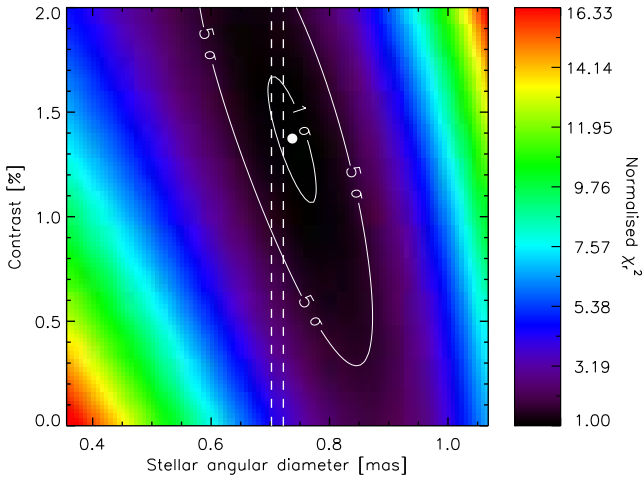}
		\caption{Normalised $\chi^2$ map for a model of an oblate photosphere (mean limb-darkened angular diameter in horizontal axis) surrounded by a uniform disc filling the whole field-of-view (in relative flux, vertical axis). The diameter predicted by the surface-brightness relations is represented by the vertical lines and the position of the best-fit model by a white filled circle.}
		\label{fig:chi2map}
	\end{flushleft}
\end{figure}
Using the approach described above, we produced the normalised $\chi^2$ map shown in Fig.~\ref{fig:chi2map} as a function of the mean\footnote{The mean angular diameter corresponds to the geometric mean of the minor and major axes of the elliptical photosphere.} angular diameter of the photosphere and of the H-band disc/star contrast. The two contours at 1 and 5-$\sigma$ indicate the parameter space that falls within the respective confidence levels taking the number of degrees of freedom in the $\chi^2$ distribution into account. In the present case, we assume that all spectral channels are fully correlated so that there are 21$\times$6 independent measurements and two parameters to fit, hence 124 degrees of freedom. Using the surface brightness relations \citep[SBR,][]{Kervella:2004d} over various wavelength bands (B, V, J, H, K), we derived a stellar angular diameter of $\theta_{\rm LD}= 0.712 \pm 0.010$\,mas. This value does not significantly depend on the chosen photometric bands, which is a good sign of robustness.
%an effective temperature of 8200~K, a $\log g$ of 4.1, and [M/H]=0.05,
%%--
For a purely photospheric model, the best-fit diameter is 0.875 $\pm$ 0.011 $\pm$ 0.012~mas, where the second and the third terms represent the statistical and the systematic errors, respectively. This value is at more than 10-$\sigma$ from the one given by the SBR. Moreover, the $\mathcal{V}^2$ residuals show a systematic trend versus spatial frequencies (Fig.~\ref{fig:globalfit}, bottom panel). Now accounting for a possible excess emission, the best-fit model is obtained for a stellar diameter of 0.736 $\pm$ 0.015 $\pm$ 0.012~mas and a disc/star contrast of 1.37 $\pm$ 0.10 $\pm$ 0.13\%. The best-fit diameter is fully compatible with the one given by the SBR, and no other trend is seen in the residuals (Fig.~\ref{fig:globalfit}, middle panel). We conclude that our dataset unambiguously demonstrates the existence of an H-band circumstellar emission around \object{$\beta$~Pic} and in the field-of-view of PIONIER. This excess emission is repeatedly detected in each run with at least a 5-$\sigma$ confidence level and with compatible values within error bars. Using our best-fit angular diameter, we derived best-fit flux ratios of 1.48 $\pm$ 0.13 $\pm$ 0.17\% and 1.32 $\pm$ 0.10 $\pm$ 0.13\% by fitting the data of observing runs A\&B and C separately. This suggests that the excess emission was present and did not vary during the time covered by our observations, i.e. approximately one year.

\begin{figure}[!t]
\centering
\includegraphics[height=6.4 cm]{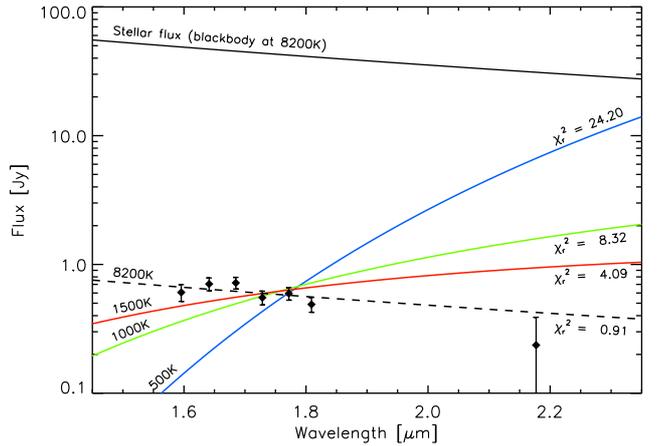}
\caption{Best-fit resolved fluxes given as a function of wavelength, assuming that \object{$\beta$~Pic} emits as a blackbody of temperature 8200\,K. The solid lines give the flux expected from blackbody sources of various equilibrium temperatures and the optimum emitting surface area (computed by $\chi^2$ minimisation). The dashed line represents the flux from a blackbody source of 8200~K or, equivalently, from scattered stellar light ($\chi_r^2$=0.91). The K-band flux obtained from archival VLTI/VINCI measurements is shown for comparison but is not included in the fit (see more information in the main text).}\label{fig:fit_flux}
\end{figure}

Finally, we revisited the K-band $\mathcal{V}^2$ measurement obtained in 2002 with the VLTI/VINCI instrument. These observations used a single and relatively long baseline length ($\sim$92\,m), making it impossible to lift the degeneracy between the stellar diameter and a circumstellar emission \citep[see Fig.~7 in][]{difolco:2004}. By using our best-fit stellar angular diameter (i.e., 0.736 $\pm$ 0.015 $\pm$ 0.012~mas), we derived a best-fit K-band flux ratio of 0.76\% $\pm$ 0.49\%, which will be used in the following discussion.

\section{Constraining the resolved emission}\label{sec:modelling}

To constrain the origin of the resolved near-infrared emission, we first tried to fit the data with disc models of various geometries \citep[see similar analyses in ][]{Absil:2009,Defrere:2011}. This analysis provided only a constraint on the minimum spatial extent of the near-infrared excess emission zone, which must be resolved by PIONIER on all baseline lengths and orientations (i.e., further than $\sim$0.1\,AU from the star). Then, we fitted the data corresponding to each spectral channel separately, considering the procedure described in the previous section and the stellar angular diameter derived with all data (i.e., 0.736 $\pm$ 0.015 $\pm$ 0.012~mas). Given the low spectral resolution of PIONIER (R$\approx$40), we assumed that \object{$\beta$~Pic} emits as a blackbody of temperature 8200\,K and converted the dispersed best-fit contrasts into fluxes. The best-fit fluxes were then fitted to pure blackbody models of various temperatures as represented in Fig.~\ref{fig:fit_flux}. This figure shows that a relatively high temperature ($\gtrsim$1500\,K) is required to fit the data in a satisfactory way ($\chi^2_r<5$). Including the K-band best-fit flux in the fit would reinforce this statement, although this must be considered with care given the eight to nine years separation and the greater uncertainty of the K-band measurements. Models corresponding to temperatures in the range of 1500-2000\,K would suggest the presence of hot material in the field-of-view of PIONIER while higher temperature models would rather point towards the scattering of stellar light by circumstellar dust ($\sim$8200~K, $\chi_r^2$=0.91) or towards a binary companion ($\gtrsim$2000~K). The latter scenario can, however, be excluded with good confidence based on other studies \citep{Absil:2010,Lagrange:2012b} so it is not addressed in the following discussion.

%\begin{figure}[!t]
%\centering
%\includegraphics[height=6.8 cm]{beta_pic_chi2map_bb.eps}
%\caption{Reduced $\chi^2$ map obtained by fitting our dispersed H-band measurements to the emission of blackbody sources of various temperatures and emitting surfaces.}\label{fig:fit_map}
%\end{figure}

\section{Discussion}

The most attractive scenario for explaining the wavelength-independent flux ratio across the H band is the scattering of stellar light in the outer debris disc ($\chi^2_r=0.91$, see Fig.~\ref{fig:fit_flux}). Since the disc is seen edge-on, a significant amount of stellar light could be forward- (or back-) scattered towards the interferometer field-of-view. To assess the validity of this scenario, we performed radiative transfer modelling with the \textit{GraTer} software \citep{Augereau:1999}. This software computes synthetic fluxes and images of dust discs in scattered light and thermal emission over various parameters for the dust composition and for the spatial and grain size distributions. It accounts for the spatial filtering of the interferometers and uses the Mie theory to  self-consistently simulate anisotropic scattering depending on grain properties. Considering a wide range for the dust parameters, we produced disc models that fit well our dispersed H-band best-fit flux ratios, the K-band contrast computed from archival VLTI/VINCI measurements, and upper limits in the mid-infrared provided by single-dish telescopes \citep{Telesco:2005}. The results show that light scattered in the known debris disc might account for a significant part (up to 70\%) of the near-infrared excess flux, while producing a very good match to archival mid- to far-IR data through its thermal emission. This maximum value of 70\% must be considered as very conservative since the Mie theory tends to overestimate forward- and back-scattering owing to the spherical grain hypothesis. The spectral slope in the near-IR is reproduced well by this scenario, but additional hot circumstellar material must be present in the vicinity of \object{$\beta$~Pic} in order to account for the remainder of the measured H-band flux ratio.

The presence of hot material around \object{$\beta$~Pic} would not be a surprise and might have various origins, such as stellar wind (free-free emission), mass-loss events, hot gas, or hot exozodiacal dust. Stellar winds and mass-loss events can, however, be excluded with very good confidence since they are expected to be very weak in the case of A-type stars \citep[see discussion in][]{Absil:2008b}. Hot gas is likely to be present in the inner system \citep[$\lesssim$ a few AU,][]{Hobbs:1988} and must emit in the continuum in order to account for our measurements. Assuming that the gas is produced by evaporating comets and braked by a ring of neutral hydrogen located at about 0.5~AU, \citet[][]{Lagrange:1998} derived temperature estimates of 1500-2000\,K with a gas production rate of $\sim$$10^{-16}$ M$_\odot$/yr. The question is thus whether hot gas could accumulate in sufficient quantities to contribute (at least partially) to the detected near-infrared emission. This is, however, beyond the scope of this paper and will not be addressed here. A more likely scenario is the presence of hot exozodiacal dust as proposed for older A-type stars \citep[e.g.,][ Mennesson et al., submitted to ApJ]{Absil:2006b,Defrere:2011}. In the case of \object{$\beta$~Pic}, the hot dust in the inner region of the disc could be produced by comets (``falling evaporating bodies'') dynamically perturbed by one or several planets \citep[][Bonsor et al., submitted to A\&A]{Beust:1990}. More investigations are clearly needed to establish the most realistic scenario and, particularly, the balance between hot material and scattered light. This will be the subject of a forthcoming paper (Lebreton et al., in prep) that will include both near-infrared (H and K bands, this work) and mid-infrared (di Folco et al., in prep.) resolved observations.

%Unveiling the origin of such a population of hot material may bring more insight on the surrounding planetary system.

\begin{acknowledgements}
DD, JBLB, JL, OA, JCA, and SE thank the French National Research Agency (ANR) for financial support through contract ANR-2010 BLAN-0505-01 (EXOZODI). DD acknowledges the support of EII (Fizeau programme). OA acknowledges the support from an F.R.S.-FNRS Postdoctoral Fellowship. PIONIER was originally funded by the Poles TUNES and SMING of Universit\'e Joseph Fourier (Grenoble) and subsequently supported by INSU-PNP and INSU-PNPS. The integrated optics beam combiner is the result of collaboration between IPAG and CEA-LETI based on CNES R\&T funding. The authors thank all the people involved in the VLTI project. We used the Smithsonian/NASA Astrophysics Data System (ADS) and the ``Centre de Donn\'ees astronomiques de Strasbourg'' (CDS).
\end{acknowledgements}

%\vspace{-0.5cm}

\end{document}